\documentstyle[cite,aps,epsfig]{revtex}
\input epsf
\input rotate

\advance\topmargin by 0.5cm
\advance\oddsidemargin by 1cm
\advance\evensidemargin by 1cm

\begin{document}
\newcommand{\lamtil}{{\tilde\lambda}}
\newcommand{\ktil}{{\tilde k}}
\newcommand{\dbydrho}{{\partial_\rho}}
\newcommand{\dbydrhot}{{\partial_\rho^2}}
\newcommand{\dbyds}{{\partial_s}}
\newcommand{\dbydst}{{\partial_s^2}}
\newcommand{\nvec}{{\bf n}}
\newcommand{\tvec}{{\bf t}}
\newcommand{\imag}{{\rm i}\,}
\newcommand{\tilomr}{\tilde\omega_r}

\begin{center}
\Large\sc
Pattern formation in directional solidification under shear flow.
I: Linear stability analysis and basic patterns.
\end{center}

\begin{center}
\large
Yannick Marietti$^{1}$, Jean-Marc Debierre$^{1}$,
Thomas M. Bock$^{2}$, Klaus Kassner$^{3}$
\vskip 0.2 cm
$^{1}$ Laboratoire Mat\'eriaux et Micro\'electronique de Provence, Case 151,
Facult\'e des Sciences de St.~J\'er\^ome, 13397 Marseille C\'edex 20\\
$^2$Institut f\"ur Experimentalphysik V, Universit\"at
Bayreuth, D-95440 Bayreuth,  Germany\\
$^3$Institut f\"ur Theoretische Physik,
Otto-von-Guericke-Universit\"at Magdeburg
Postfach 4120,
D-39016 Magdeburg, Germany\\
E-mail: Klaus.Kassner@Physik.Uni-Magdeburg.de\\

\vskip 0.2 cm
\vskip 0.2 cm
December 12, 2000
\\[0.5cm]


\end{center}


\vskip 1cm

\noindent
Abstract\\

An asymptotic interface equation for directional solidification
near the absolute stabiliy limit is extended by a nonlocal term describing
a shear flow parallel to the interface. In the long-wave limit considered,
the flow acts destabilizing on a planar interface. 
Moreover, linear stability analysis suggests that the  morphology diagram
is modified by the flow near onset of the Mullins-Sekerka instability.
Via numerical analysis, the bifurcation structure of the system
is shown to change.
Besides the known hexagonal cells,
structures consisting of
stripes arise. 
Due to its symmetry-breaking
properties, the flow term induces a lateral drift of the whole pattern,
once the instability has become active. The drift velocity is measured
numerically and described analytically in the framework of a linear
analysis. At large flow strength, the linear description breaks down,
which is accompanied by a transition to flow-dominated morphologies, described
in a companion paper.
Small and intermediate flows lead to increased order in the lattice
structure of the pattern, facilitating the elimination of defects. 
Locally oscillating structures appear closer to the instability threshold 
with flow
than without.
\\



\newpage

\newcommand{\beq}{\begin{equation}}
\newcommand{\beqa}{\begin{eqnarray}}
\newcommand{\eeq}{\end{equation}}
\newcommand{\eeqa}{\end{eqnarray}}
\newcommand{\half}{\frac{1}{2}}
\newcommand{\eps}{\epsilon}
\newcommand{\kom}{\>,}
\newcommand{\pnt}{\>.}


\section{Introduction}
Directional solidification is an important experimental procedure both from the
applied and the fundamental points of view.

On the one hand,
growth techniques such as zone melting and the Bridgman method were
developed in the fifties to purify semiconductor materials.
Subsequently, the basic process became an important industrial tool,
and it continues to remain the basis of fundamental metallurgical
techniques.

On the other hand,
when an alloy is solidified by pushing its melt at a constant velocity
along an imposed thermal gradient towards lower temperatures --
this  realization of directional solidification is particularly amenable
to theoretical analysis -- the interface between the liquid and the solid,
which is flat at zero velocity, undergoes a morphological instability.
Fundamental interest in
this so-called Mullins-Sekerka (MS) instability \cite{mullins64} motivated
many current research studies of directional solidification.
The instability is driven by impurity diffusion, 
it appears at a critical speed,
and once it has set in,
new structures develop, depending on a number of factors such as the
interface roughness, the segregation coefficient, impurity diffusion, and
the impurity concentration in the liquid.
If the latter is high enough,
two-phase eutectic structures may arise \cite{jackson66},
which are {\em not} created by the MS instability.
In this article, we will rather
focus on the case of smaller concentrations leading
to the growth of a {\em dilute alloy} forming a
single-phase solid.

Experimentally, pattern formation in directional solidification has been
studied in
great detail using transparent model substances \cite{jackson65} in order
to gain fundamental insight
\cite{trivedi85,decheveigne86,decheveigne88,kurowski90,decheveigne92}.
It has also been investigated in metals, with at least a view to
applications \cite{burden74,nguyenthi97,zimmermann89,dupouy98}.
Recently, successful observations of {\em three-dimensional} directional
solidification
of transparent alloys \cite{noel97,noel98}
have opened the road to an understanding that goes beyond the description of
two-dimensional structures, for which  a large amount
of theoretical work does 
exist
\cite{caroli82,wollkind84,ungar84,dombre87,benamar88b,kessler89,brattkus90a,%
levine90,%
misbah90,classen91,kassner91d,valance92,rappel92,kassner94}.

As a consequence, it has now become an important task
to follow up with theoretical work on three-dimensional directional growth.
Detailed analysis of three-dimensional patterns has so far largely been
restricted to free growth of small structures such as single dendrites, both
analytically \cite{benamar93,brener93,brener95} and
numerically
\cite{kobayashi94,karma96,abel97}.
Three-dimensional directional solidification has been considered in the context
of phase-field modeling for small systems consisting of some ten cells
\cite{grossmann93}.
Only fairly recently, larger systems containing 
several thousand cells have been
treated within the framework of an asymptotic interface
equation\cite{kassner98}.
The main advantage of this equation is that
it reduces the dynamical problem to a description of the interface alone, which
replaces the 3D problem with a 2D one, and that, moreover, it is a local
description, thus rendering large systems tractable.
It was obtained by a Sivashinsky-type expansion about the point of absolute
stability
\cite{kassner94}, neglecting both solute trapping and deviations from local
interface equilibrium. As these approximations become doubtful for large
solidification
speeds, this model equation had to be considered a qualitative description only
and comparison with experiments had to rely  on genericity arguments.
However, we believe that the equation would be quantitatively appropriate
for the description of directional ordering experiments on liquid crystals
\cite{simon88,simon89,flesselles91} and hence we suggest the more intriguing
dynamic long-time effects predicted by it \cite{kassner98}
to be observable in liquid crystals, which also have response times more
accessible
to experiments\cite{footnote1}. In any case, statistical properties
obtained from
the simulation compared reasonably well with the same properties measured in
experiments. This holds apart from a systematic deviation allowing us to
conclude that
experimental cellular patterns do {\em not} start growing from a random
(Poissonian)
distribution.

In earth-bound experiments, convection always exerts a major influence on the
arising patterns \cite{jamgotchian99}. Therefore, we extend our previous work
here to include a flow in the description. Since this is but a first approach
to a large-scale system with flow, in order
to facilitate the subsequent analysis of results, we consider
the conceptually simplest case, an externally imposed flow parallel
to the interface, approaching a constant velocity far from it.
For this situation, representing the ``poor man's convection''
\cite{davis90},
an asymptotic description has been derived by
Hobbs and Metzener \cite{hobbs92}. One-dimensional interfaces in the presence
of flow have been studied in \cite{kassner96a}. Whereas in the latter 
investigation
solute trapping and the deviations from equilibrium of the interface were
fully taken into
into account using the Aziz model \cite{aziz82}, we will restrict ourselves
here to the considerably simpler situation of a constant 
segregation coefficient
and neglect interface kinetics, to separate generic effects from
those that are only important at high speed. Also the full equation is more
susceptible to numerical instabilities, because its solutions
tend to develop large amplitudes taking them out of the validity
domain
of the expansion \cite{kassner98}. The only difference between the
interface equation employed in \cite{kassner98} and here will therefore be
the term generated by the flow. It renders the equation nonlocal which
already makes its simulation much more tedious than that of the original
equation and restricts our simulations to a few hundred solidification cells
rather than a few thousand.

This article consists of five sections.
In Sec.~II, we give the basic equations together with
a linear stability analysis, leading to the prediction of new morphologies
induced by the flow. Section III describes the numerical approach and
our methods of velocity measurement. We give
a number of simulation  results in Sec.~IV, exhibiting the basic patterns
in the  part of parameter space where some degree of spatial order prevails.
A more detailed analysis and characterization of these structures is presented
in the subsequent article \cite{marietti00b}, henceforth referred to as II.
 Conclusions regarding flow effects on pattern dynamics and the bifurcation
structure of the system
are  summarized in Sec.~V. An appendix provides the
 real-space representation of the flow term thus displaying its
nonlocal nature.
Some analytic approximations are worked out in detail in a second appendix.

%

\section{Model equation and linear stability analysis}

\subsection{Long-wave equation}

The equation to be studied is
\newcommand{\kinv}{\frac{1}{k}}
\newcommand{\kinvt}{\frac{2}{k}}
\newcommand{\gbar}{\bar{G}}
\newcommand{\abs}[1]{\left\vert #1\right\vert}
\beqa
&&\zeta_{tt} - \left(2+\kinv+\nu\right) \nabla^2 \zeta_t
             + \left(1+\kinv+\nu^2\right) \nabla^4 \zeta
             + 8 k \nabla^2 \zeta + 8 k \gbar \zeta -f
\frac{\partial}{\partial x} {\cal L}[\zeta]
             \nonumber \\
& = & 2 \zeta_t \nabla^2 \zeta + 2\left(\abs{\nabla \zeta}^2\right)_t
    - 2 \nabla^2 \left(\abs{\nabla \zeta}^2\right)  - 2 \nu (\nabla\zeta)
\nabla \left(\nabla^2 \zeta\right)-\kinvt \nabla\left\{(\nabla\zeta)
\nabla^2 \zeta\right\}
              \nonumber \\
&&
    \mbox{}
    - 2 \nabla \left\{(\nabla\zeta) \abs{\nabla \zeta}^2\right\} \pnt
    \label{basiceq}
\eeqa
We denote by $\nabla$ the  two-dimensional gradient operator,
$\nabla = \left(\partial_x,\partial_y\right)$.
The position of the liquid-solid interface is $\zeta(x,y,t)$
and coordinates are chosen such that the temperature gradient
is oriented along the $z$ axis, whereas the flow is parallel to the $x$ axis.
${\cal L}[\zeta]$ is a linear but nonlocal
functional of $\zeta$, given via its Fourier transform
\beq
{\cal F}\Bigl[{\cal L}[\zeta]\Bigr]({\bf p}) = \vert{\bf p}\vert\> {\cal
F}[\zeta]({\bf p}) \>,
\label{ftnonloc}
\eeq
with ${\cal F}$ denoting the  transform,
${\cal F}[\zeta] = \int_{-\infty}^\infty dx \int_{-\infty}^\infty dy\,
\zeta({\bf x})
\exp(-\imag {\bf p x })$,
and ${\bf p}$ its wave vector argument.
Position space representations for ${\cal L}[\zeta]$ in one and two dimensions
are given in App.~A.

There are four nondimensional parameters in our equation:
 the segregation coefficient $k$, the  ratio  $\nu=D_s/D$ of the diffusion
 coefficients for impurities
in the solid and the liquid, the  nondimensional
temperature gradient  $\gbar$, and the strength $f$ of the flow.
The {\em one-sided
model} is characterized by  $\nu=0$,  the {\em symmetric model} by  $\nu=1$.
  $\gbar$ is related to physical parameters   via:
\beq
\gbar = \frac{8 D^3 L^2 m \Delta c}{\gamma^2 T_m^2} \,
 \frac{G}{V(V_a-V)^2} \pnt
\label{defgbar}
\eeq
In this expression,
$L$ is the latent heat (per unit volume)   of the transition, $m$ the absolute
value of the slope of the liquidus line, $\Delta c$ the  miscibility gap
  at the base temperature $T_0$
 ($T_0=T_m-m c_0$, with $T_m$ the melting temperature of the
pure substance, $c_0=c_\infty+\Delta c$, and $c_\infty$ the initial
concentration of the liquid), $\gamma$ is the surface energy,
assumed isotropic here,
$V_a$ is the velocity corresponding to the absolute stability limit,
given by $V_a = m L \Delta c D/ \gamma T_m k$.
$G$ is the temperature gradient and $V$ the pulling (or pushing) velocity.
Finally, $f$ is given in terms of dimensional quantities as
\beq
f = \frac{8 V_\infty}{\epsilon V S}\>, \label{deff}
\eeq
$V_\infty$ being the speed of the imposed flow far from
the interface, $S=\nu_k/D$ is the Schmidt number ($\nu_k$ is the kinematic
viscosity of the liquid),
and $\epsilon$ is  the nondimensional distance from absolute
stability,
$\epsilon=1/2k-\gamma T_m V/2 m L \Delta c D$.
$f$ is dynamically determined by the ratio of the flow and pulling speeds.

Equation (\ref{basiceq}) is obtained from an expansion about the absolute
stability limit, where the wavelength of the most unstable mode
diverges as $1/\sqrt{\epsilon}$. It takes the
form of a strongly nonlinear long-wave equation, in which
wavelengths have been rescaled by $\sqrt{\epsilon}$ to make them $O(1)$.
Thus, nondimensional lengths are measured in units of a rescaled
diffusion length along the $x$ and $y$ directions, with a scaling
factor $1/\sqrt{\epsilon}$; lengths in the $z$
direction are measured as multiples of the unscaled diffusion length.
The time unit is a diffusion time ($2D/V^2$) scaled by $1/\epsilon$.
For $k=0$, the equation  becomes indefinite, because there
is no absolute stability limit in this case.

For simplicity, we shall
set $k=1$ in the following, i.e., $\Delta c =$ const, independently of the
reference
temperature $T_0$.
{}From previous experience, the choice of $k$ is not expected to have a
strong influence
on results, as long as $k$ does not become very small.

To  further reduce the parameter space to be explored, we also set $\nu=1$,
thus
restricting ourselves to the symmetric model.  This is not a particularly
realistic model for directional
solidification, where $\nu=0$ would be more appropriate.
 However, for directional ordering in liquid crystals,
it is a better approximation than the one-sided model.
Moreover, the choice $\nu=1$
is least problematic in numerical simulations with a direct finite-difference
discretization, because deep grooves that may trigger numerical instabilities
are less likely to evolve with diffusion in the solid allowed. Since the most
efficient way
to deal with the nonlocal term is to work in Fourier space, we have developed
a pseudospectral code, which is much less susceptible to these problems.
Nevertheless, a comparison with previous results \cite{kassner98} is easier,
if we keep $\nu$ at the value then used, and experience suggests
\cite{kassner94}
that generic patterns are not strongly influenced by this choice.
Finally, because
our results must not be expected to be quantitative except for liquid
crystal systems,  we stick to the  value $\nu=1$ here. Tests with various
nonzero values of $\nu$ have been performed but
have
not revealed interesting differences.

Therefore, the important parameters to be varied in the simulations are
$\gbar$ and $f$.

The form of Eq.~(\ref{basiceq}) can be guessed from symmetry 
and scaling arguments.
The linear terms on the left hand side are determined (including their
coefficients) by the
linear stability analysis of the full three-dimensional model, involving
diffusive transport
and the coupling to the Navier-Stokes equations. Scaling arguments tell us that
the nonlinear terms can contain at most four spatial derivatives and for each
temporal derivative present there must be two spatial derivatives less
(since wavenumbers scale as $\sqrt{\epsilon}$ but frequencies as
$\epsilon$).  In the absence of a thermal gradient, we have
translational symmetry in the $z$ direction, so we
know that all nonlinear terms must contain derivatives of $\zeta$ only. If
there is no flow, we also have parity symmetry, which constrains the number of
spatial derivatives of
terms not containing $f$
to being even. Finally, rotational symmetry can be invoked
to exclude terms such as $(\nabla\zeta)^4$ \cite{kassner94}. From these
considerations,
one obtains all the nonlinear terms on the r.h.s., but not their prefactors,
of course, for which the full expansion must be performed
\cite{hobbs92,kassner96}.
What can also be guessed is that the flow should lead to a nonlocal term
 breaking
the parity symmetry with respect to the $x$ coordinate.
It is not clear beforehand, however,
that it does {\em not} introduce additional nonlinear {\em local} terms.
The nonlocal term on
the left-hand side is, in a sense,
the simplest nonlocality possible (see Appendix A).

\subsection{Linear stability analysis}

The problem of coupled morphological and convective instabilities has
a long history of detailed study
\cite{coriell80,hurle82,hurle83,coriell84,davis90}.
In the case of buoyancy-driven convection with a lighter solute and
solidification proceeding upward in the gravitational field (leading
to unstable stratification), it was found that in most cases the coupling
between the instabilities is weak due to a large disparity in
unstable wavelengths \cite{caroli85}, even though an oscillatory instability
may occur \cite{riley89} in special circumstances.
For Rayleigh numbers below the critical value,
convection delays the Mullins-Sekerka
instability in the limit of small segregation
coefficients, where  a long-wave equation different from ours can be derived.
Forced flows were studied by Coriell {\em et al.}
\cite{coriell84} and by Forth and Wheeler \cite{forth89},
who found that for two-dimensional disturbances the flow delays the
morphological instability, whereas for  disturbances
with wave vectors perpendicular to the flow the latter
normally does not affect the critical conditions of the MS instability.
However, for small-wavenumber modes, the MS instability can be enhanced
\cite{forth89}. Near absolute stability, the wavelength is much larger than
the diffusion length and parallel flow was shown to be destabilizing
to disturbances that travel against
it  \cite{hobbs91}. This is the situation encountered here.

Equation (\ref{basiceq}) has  the steady-state solution $\zeta\equiv 0$.
Obviously,
the linear stability analysis of this solution will involve only the terms
on the left-hand side of the equation. Inserting the perturbation ansatz
$[{\bf x} = (x,y)]$
\beq
\zeta = \zeta_1 \exp\left(\omega t + \imag {\bf q} {\bf x}\right)
\label{linansatz}
\eeq
into Eq.~(\ref{basiceq}), we obtain the dispersion relation
($q=\vert{\bf q}\vert$)
\beq
\omega^2 + \left(2+\frac1k + \nu\right) \omega q^2 +
\left(1+\frac1k + \nu^2\right)q^4
-8 k q^2 + 8 k \gbar -\imag f q_x q = 0 \>.
\label{compldisprel}
\eeq
The terms stemming from the partial derivatives are obtained in a
straightforward
manner, only the one arising from the
nonlocal term may require some explanation.
To compute the nonlocal term for the perturbation (\ref{linansatz}) we
first take the spatial Fourier transform of $\zeta$, which is simply
$ 4\pi^2 \zeta_1 \exp(\omega t) \delta\left({\bf q} + {\bf p}\right) $,
then multiply it by $\vert {\bf p}\vert$ to obtain the transform of ${\cal
L}[\zeta]$.
Transforming back we get $\zeta_1 \vert {\bf q} \vert \exp(\omega t 
+ \imag {\bf q} {\bf x})$,
the derivative of which with respect to $x$ produces a prefactor $\imag q_x$.
After dropping
the common exponential factor and the prefactor $\zeta_1$ of all terms, we
are left
with the $q$ dependent expression of Eq.~(\ref{compldisprel}).

Setting $\omega = \omega_r + \imag \omega_i$ and
decomposing the dispersion relation into its real and imaginary parts,
 we obtain
\beqa
\omega_r^2 -\omega_i^2
+ \left(2+\frac1k + \nu\right) \omega_r q^2 + \left(1+\frac1k + \nu^2\right)q^4
-8 k q^2 + 8 k \gbar &=& 0 \>, \label{realdisprel} \\
\omega_i\left[2\omega_r+\left(2+\frac1k + \nu\right) q^2 \right] &=& f q_x
q \>.
\label{imagdisprel}
\eeqa
The unstable mode takes the form
\beq
\zeta = \zeta_1 \exp\left\{\omega_r t+\imag \left[q_x
\left(x+\frac{\omega_i}{q_x} t\right)+q_y y\right]\right\} \>,
\label{travelwav}
\eeq
i.e., it corresponds to a traveling wave
moving along the $x$ axis at velocity $V_d = -\omega_i/q_x$.
On the neutral surface, $\omega_r=0$, which implies
\beq
\omega_i = \frac{f q_x}{\left(2+1/k + \nu\right)q} \>,
\label{omegaineutral}
\eeq
hence we have a Hopf bifurcation whenever the flow is different from zero
and the pattern is oriented such that $q_x\ne0$.
The  velocity of the corresponding traveling wave is simply
\beq
V_d = -\frac{f}{\left(2+1/k + \nu\right)q} \>.
\label{vdrift}
\eeq

\begin{figure}[h]
\centerline{ \epsfig{file=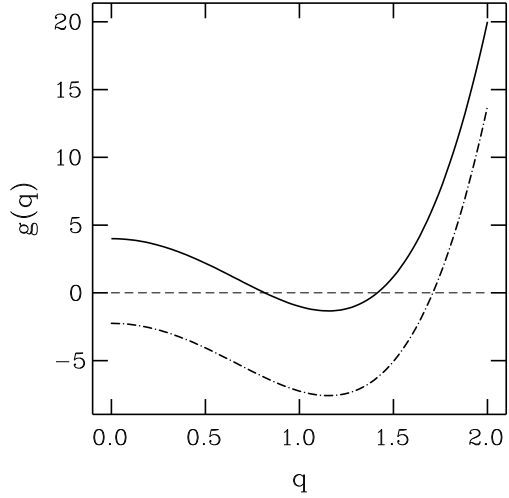,width=8.5cm,angle=0} }
\caption{The function $g(q)$ determining the neutral surface. $\gbar=0.5$.
Upper curve: $f=0$, lower curve: $f=10$.}
\label{gofqpict}
\end{figure}

Inserting (\ref{omegaineutral}) into (\ref{realdisprel}), we arrive at the
equation for
the neutral surface:
\beq
g(q) \equiv \left(1+\frac1k + \nu^2\right)q^4 - 8 k q^2 + 8 k \gbar -
\left(\frac{f q_x}{\left(2+1/k + \nu\right)q}\right)^2 = 0 \>,
\label{neutralcurve}
\eeq
where the last term depends on the angle between the flow and the wave
vector only,
not on the modulus of the latter, a situation that we describe by setting
$\cos \phi = q_x/q$.
Obviously, for fixed values of $f$ and $\phi$
the essential change of the neutral curve
(in the $q\gbar$ plane) brought about by the flow is an increase of
the critical value of $\gbar$ where the instability first appears.
Hence  the flow has a {\em destabilizing} effect,
since the region of parameter space, where the planar solution is unstable
corresponds to values of $\gbar$ {\em below} the critical value.
For given $\gbar$, the flow can be made large enough to render the planar front
unstable even near $q=0$, i.e., with respect to homogeneous perturbations.
Figure~\ref{gofqpict} displays the function
$g(q)$ for two values of the flow (assuming $\cos\phi=1$). With weak flow,
 $g(q)$ is positive at $q=0$, and if $\gbar$ is small enough, the function
has two
zeros at positive $q$, i.e., a band of modes containing only finite $q$
values is
unstable. For strong flow,  $g(q)$ is negative  at $q=0$, i.e., {\em all} modes
up to the marginal one 
 are unstable.

\begin{figure}[h]
\centerline{ \epsfig{file=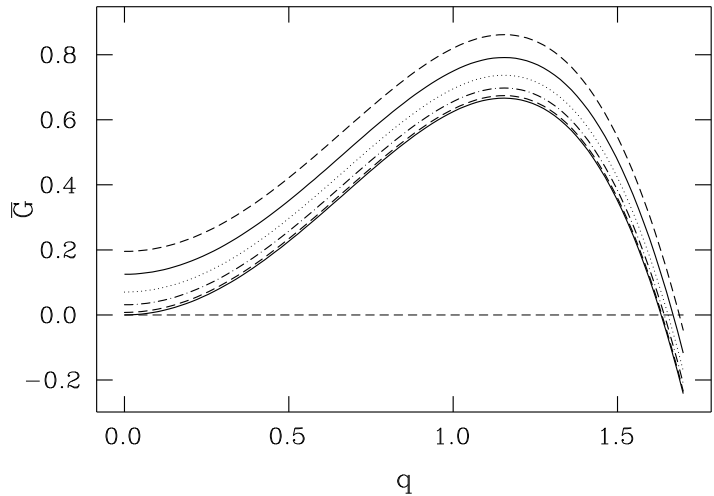,width=10.0cm,angle=0} }
\caption{Neutral curve for different flow strengths $f$. Lower solid line:
$f=0$, dashed line: $f=1$, dash-dotted line: $f=2$, dotted line: $f=3$,
solid line: $f=4$, higher dashed line: $f=5$. The curves shift upward with
increasing $f$ by an amount proportional to $f^2$.}
\label{neutral_curve}
\end{figure}

The neutral surface is the set of zeros of $g(q)$ in the  space spanned by
$q$, $\gbar$, and $f$. For convenience, we restrict ourselves to the
$q\,\gbar$ plane
and draw neutral curves for several discrete values of the flow parameter
$f$ in
Fig.~\ref{neutral_curve}, to demonstrate  both the $\gbar$ and $f$ dependences
of the neutral surface.

A calculation of the
critical value of the temperature
gradient in the general case (requiring $d\omega_r/dq=0$) leads to
\beq
\gbar_c = \frac{2k^2}{1+k + k\nu^2} +
\frac{f^2 k \cos^2\phi}{8\left(1+ 2k + k \nu\right)^2}\>,
\label{threshold}
\eeq
exhibiting the fact that the instability threshold depends on the angle between
the flow and the wave vector of the perturbation. Without flow, the threshold
is given by the first term of Eq.~(\ref{threshold}) \cite{kassner94}
and the bifurcation, transcritical
in two dimensions, is known to give rise to hexagonal patterns at onset
\cite{daumont97}.
{}From the equation,
we can immediately conclude that for values of the
temperature gradient satisfying $ \gbar> \gbar_c(f=0)$, there exists
a critical flow strength given by
\beq
f^2 = \frac{8(1+2k+k\nu)^2}{k}
 \left(\gbar- \frac{2k^2}{1+k + k\nu^2}\right) \>,
\eeq
above which the planar front is destabilized by the flow alone.
In this case, the
patterns emerging as the planar interface becomes unstable
will not be hexagons but rather stripes
oriented orthogonally to the flow
(since these are the most unstable disturbances). They should drift
against the flow
with a speed approximately given by Eq.~(\ref{vdrift}) and calculated more
precisely below. As we shall see in Sec.~IV, these
predictions are borne out by the simulations.
It is then an interesting question, how the system will behave on decrease
of $\gbar$ below the zero-flow threshold.
What happens when the pattern amplitude approaches saturation cannot
be predicted from the linear analysis but 
will be discussed in some detail in II.

To get an idea of the behavior of the drift speeds to be expected
beyond the bifurcation point, we compute
the fastest-growing unstable mode, which should provide a decent
approximation to
observed wavelengths
close to threshold. For simplicity, this calculation is restricted to the
symmetric
model ($\nu=1$) with $k=1$.
The $q$ value, at which the growth
rate is maximum  is obtained by differentiating Eqs.~(\ref{realdisprel}),
(\ref{imagdisprel}) and setting
 $d\omega_r/dq=0$, which gives us two more relations
\beqa
-2 \omega_i \frac{d\omega_i}{dq} + 8q \omega_r + 12 q^3 -16 q & = & 0 \>,
\label{relmax1} \\
8 q \omega_i + \frac{d\omega_i}{dq}\left(2\omega_r+4q^2\right) &=& 2fq_x
\>, \label{relmax2}
\eeqa
from which the four unknowns $\omega_r$, $\omega_i$, $d\omega_i/dq$ and $q$
can be
determined. Two simplifications are straightforward, giving expressions for
 $\omega_i$ and $d\omega_i/dq$:
\beqa
\frac{d\omega_i}{dq} &=&
\frac{1}{2\omega_r+4q^2}\left(2fq_x-8q\omega_i\right) \>, \label{domegai} \\
\omega_i  &=& \frac{fq_x q}{2\omega_r+4q^2} \>, \label{omegai}
\eeqa
which leaves us with two equations for $\omega_r$ and $q$.

Incidentally, we can immediately gather an interesting consequence 
from these equations
regarding the question of {\em convective} versus {\em absolute} instability 
\cite{footabs}.
If we require  $\omega_r=0$ in Eq.~(\ref{omegai}), this implies 
$d\omega_i/dq=0$ by virtue of Eq.~(\ref{domegai}). Hence, at the critical
point of the linear instability, we have $d\omega/dq=0$, {\em i.e.}, the 
group velocity of a localized perturbation vanishes. This means 
the thresholds for convective and absolute instabilities coincide in
our system, which therefore is never only convectively unstable.
This statement remains true for arbitrary values of $k$ and $\nu$.

\begin{figure}[h]
\centerline{ \epsfig{file=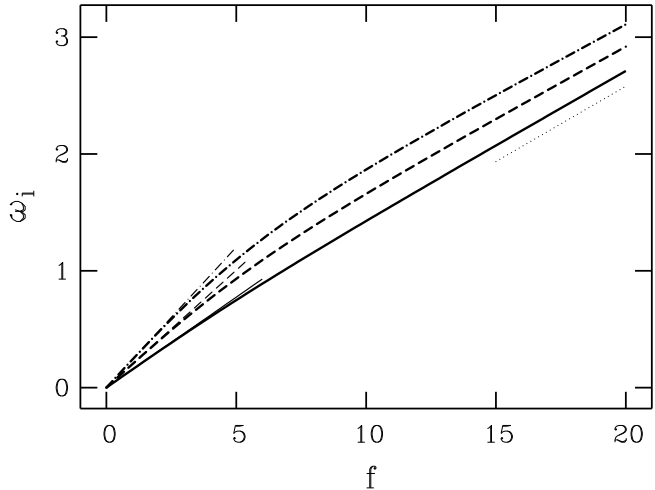,width=10.0cm,angle=0} }
\caption{Imaginary part $\omega_i$ of the complex growth rate $\omega$ for
the fastest-growing
mode (``drift frequency'') as a function of the flow parameter $f$
for $\gbar=0.1$ (solid line), $\gbar=0.35$
(dashed line), and $\gbar=0.6$ (dash-dotted line).
The thin dotted line is the asymptotic approximation for large $f$,
Eq.~(\protect\ref{omi_asympt_largef}), whereas the other thin lines denote asymptotic
expressions for small $f$ and different $\gbar$ values, 
Eq.~(\protect\ref{omi_asympt_smallf}).
}
\label{omega_i}
\end{figure}

In the following, we will assume
that the stripe pattern is oriented orthogonally to the flow
(as it usually is if it arises spontaneously from a random initial
condition), therefore $q_x=q$.
The equations determining the two remaining unknowns are then
\beqa
\omega_r^2 - \frac{f^2 q^4}{4(\omega_r+2q^2)^2} + 4\omega_r q^2 + 3
q^4-8q^2+8\gbar &=& 0 \>,
\label{eqomegar}\\
\frac{f^2 q^2 \omega_r}{2(\omega_r+2q^2)^3} - 4\omega_r -6q^2 +8 &=& 0 \>.
\label{eqq}
\eeqa
The limiting cases $f\ll1$ and $f\gg1$ of this system of equations can be
treated analytically,
detailed expressions are given
 in App.~B. For small flow velocities the interface drifts at
a speed that is proportional to $f$, whereas for large velocities,
the drift speed becomes
proportional to $\sqrt{f}$. The imaginary part of the growth rate $\omega_i$
is proportional to $f$ in both cases but with different proportionality
constants.
A numerical solution of the
system (\ref{eqomegar}), (\ref{eqq}) is not difficult.
The resulting ``drift frequencies'' $\omega_i$ are given in Fig.~\ref{omega_i}
for several pertinent temperature gradients, together with
the asymptotic analytic expressions from App.~B.
We will compare theoretical with measured drift velocities in Sec.~IV.

\section{Numerical approach}

\subsection{Discretization}

Equation (\ref{basiceq}) was simulated with periodic boundary conditions on
quadratic
grids of sizes between 32$\times$32 and 512$\times$512.
The mesh size $h$ was usually 0.5, although at thermal gradients below
0.35, we had to reduce it to keep the code stable.
With a 128$\times$128 lattice, hexagonal structures would contain on
the order of 100 cells for $h=0.5$.
Previous simulations \cite{kassner98}
have shown that this is roughly the size of a typical
dynamical grain, inside which a system at not
too low a temperature gradient manages
to get rid of all its topological
defects and to attain complete hexagonal order.
In order to have several grains in the numerical box, a number
of 256$\times$256 systems were simulated.

Temporal discretization was done by a simple explicit first-order Euler scheme.
Two variants of the code were implemented; in the first, spatial
derivatives were
approximated by second-order accurate symmetrical stencils,
whereas in the second,
a pseudospectral approach, derivatives were computed via fast Fourier
transform, i.e., they
were accurate to order $h^N$ for a mesh size $h$ and
linear grid dimension $N$. The flow
term was always evaluated via its Fourier representation, as a real-space
calculation
would have required the computation of a double integral on the whole system 
at each
lattice point (see App.~A). We will refer to the first, less accurate approach
as the mixed code and to the second as the (pseudo)spectral one.

The mixed code was useful for the treatment of larger systems
(256$\times$256) over longer
times, since it was faster than the spectral code by a factor of about six
in this case.
However, the simulation of systems with smaller temperature gradients
($\gbar\le 0.35$) necessitated the use of the spectral code, both for
accuracy and
stability reasons. In our previous study of three-dimensional rapid
solification \cite{kassner98},
gradients below 0.35 remained essentially inaccessible due to numerical
instabilities
arising at a mesh size of 0.5. Reduction of the
 mesh size mitigated the problem,
 but restricted accessible system sizes.

\subsection{Velocity measurement}

In the presence of flow, patterns  move laterally, so it became desirable to
measure their velocity. This was done via a correlation function method as
follows.
The  dynamic evolution of the interface was simulated over a time interval
extending
from $t$ to $t+\Delta t$. Then the quantity
\beq
c(\Delta x,\Delta y,\Delta t,t) =
\left\langle \left[\zeta(x+\Delta x,y+\Delta y,t+\Delta t) -
\zeta(x,y,t)\right]^2 \right\rangle
\label{corrfunc}
\eeq
was evaluated for a number of values $\Delta x$ and $\Delta y$ that were
small multiples
of the grid spacing $h$. Angular brackets denote spatial averaging (over
the whole
grid). Next the minimum of the correlation function $c$ was determined via
parabolic
interpolation from the surroundings of the minimum value obtained within
the discrete set $\{\Delta x,\Delta y\}$. An approximation to the velocity
at time $t+\Delta t$ was then obtained as $v=(v_x,v_y)$ with
$v_x=\Delta x^\ast/\Delta t$ and $v_y=\Delta y^\ast/\Delta t$,
where $\Delta x^\ast$ and $\Delta y^\ast$ were the coordinates of the minimum.

We tested this procedure on a variety of analytically prescribed interfaces.
It turned out highly reliable and accurate (in the ppm range and better) 
whenever
the interface had a constant shape, its only dynamics being a lateral drift
motion,
and the time step $\Delta t$ was not chosen too short. The accuracy
 deteriorated to fall into
the percent range, when shape changes were allowed. This is understandable,
as with
a shape-changing interface the drift velocity is not even precisely defined.
A one-dimensional example will clarify this point.
Consider
\beq
\zeta(x,t) = \sin k (x-vt) \cos \omega t \>, \label{onedexample1}
\eeq
where intuitively one would associate the velocity $v$ with the
motion of the pattern.
But we also have
\beq
\zeta(x,t) = \frac12 \sin k \left[x-\left(v-\frac\omega{k}\right)t\right]
+  \frac12 \sin k \left[x-\left(v+\frac\omega{k}\right)t\right] \>,
\label{onedexample2}
\eeq
that is, the pattern is decomposable into two waves drifting at different
velocities
$v-\omega/k$ and $v+\omega/k$.

In this case, the correlation function can be calculated analytically:
\beq
c(\Delta x,\Delta t,t) = \frac12\left[\cos^2\omega t +\cos^2\omega
(t+\Delta t)\right]
-\cos k (\Delta x -v \Delta t)  \cos \omega t \cos\omega (t+\Delta t) \>,
\label{corr1D}
\eeq
the spatial minimum of which is, for $\cos \omega t \,\cos\omega (t+\Delta
t)>0$,
 given by
$k(\Delta x -v \Delta t) = 2 n\pi$
($n=0,\pm 1,\pm2, \ldots$). The example shows that for regular structures,
$\Delta x$ (as well as
 $\Delta y$) has to be kept smaller than a wavelength in order to avoid
solutions
with $n\ne 0$. Assuming $n=0$, we find $\Delta x/\Delta t=v$. This means that
one obtains the intuitively expected result $v$ for the velocity,
 whenever $\cos\omega\tilde t$ does not change sign in the
interval $\tilde t \in [t,t+\Delta t]$ (in these -- rare -- instances, the
algorithm
will yield $\Delta x/\Delta t=v\pm\pi/k\Delta t$, i.e., the ``velocity signal''
will display peaks at (twice) the frequency $\omega$). 
We conclude that in general the algorithm is
reliable and robust.

\section{Simulation results}

\subsection{Some basic patterns}

\begin{figure}[h]
\centerline{ \epsfig{file=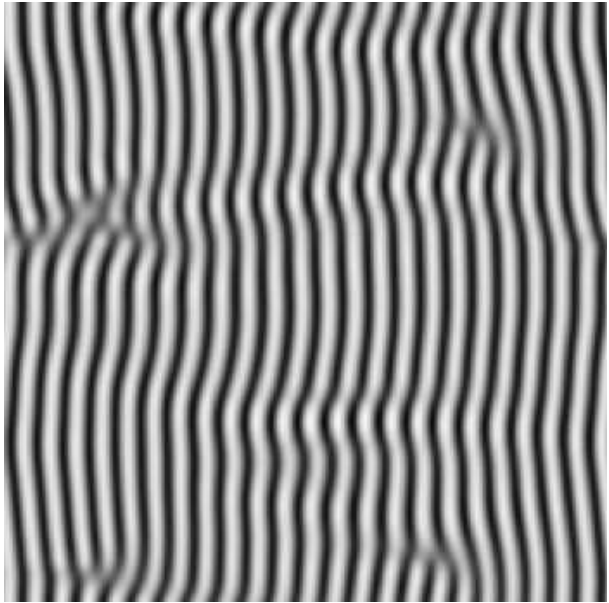,width=8.0cm,angle=0} }
\caption{Pattern at $\gbar=0.7$, $f=4.0$, system size 128.0$\times$128.0
(grid spacing $h=0.5$), i.e., the lattice is 256$\times$256. $t=150.0$.
Lengths are given in units of the
(rescaled) diffusion length, times in units of the (rescaled)
diffusion time.
}
\label{g07f04}
\end{figure}

\begin{figure}[h]
\centerline{ \epsfig{file=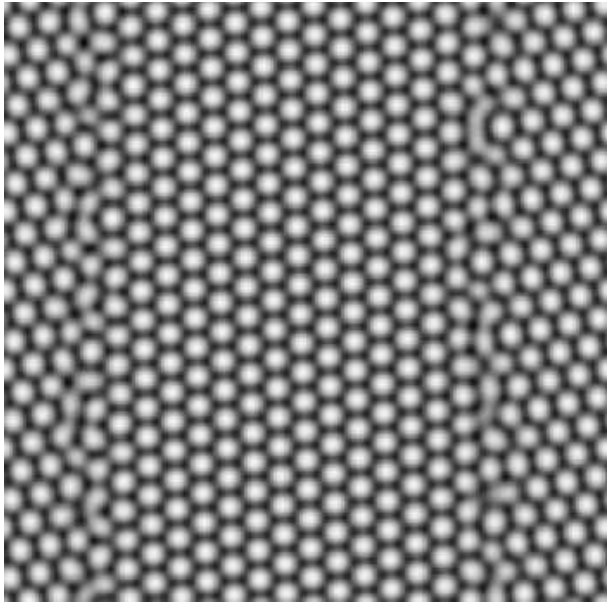,width=8.0cm,angle=0} }
\caption{Pattern at $\gbar=0.6$, $f=2.0$, system size 128.0$\times$128.0
(grid spacing $h=0.5$). $t=2000.0$.
}
\label{g06f02}
\end{figure}

\begin{figure}[h]
\centerline{ \epsfig{file=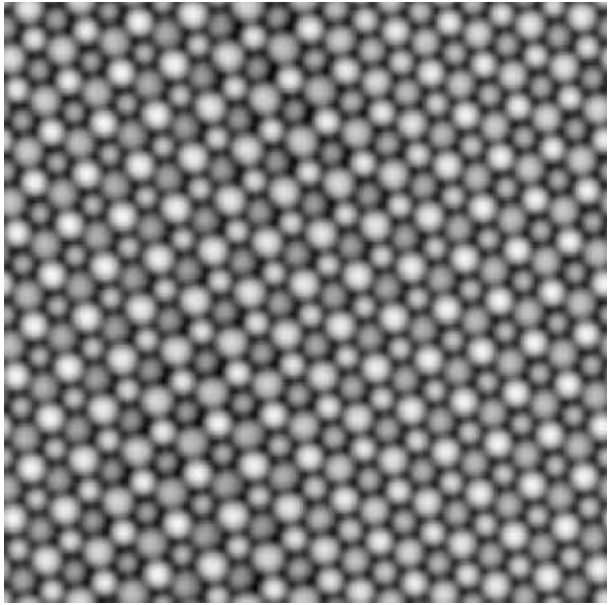,width=8.0cm,angle=0} }
\caption{Pattern at $\gbar=0.35$, $f=2.0$, system size 128.0$\times$128.0
(grid spacing $h=0.5$). $t=2000.0$.
}
\label{g035f02}
\end{figure}

\begin{figure}[h]
\centerline{ \epsfig{file=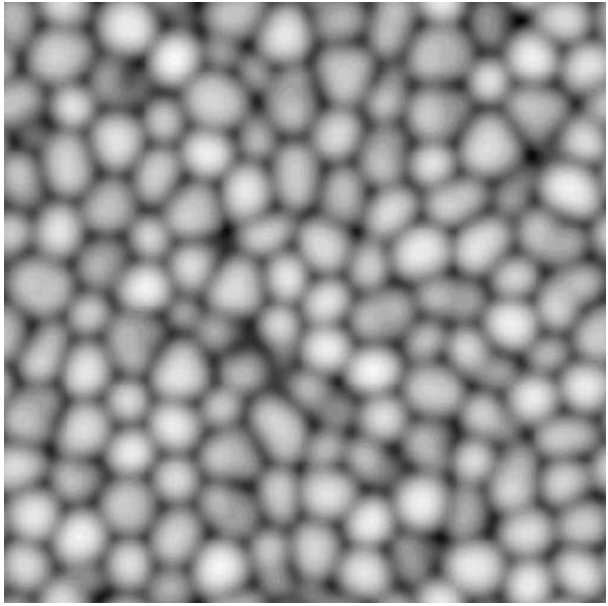,width=8.0cm,angle=0} }
\caption{Pattern at $\gbar=0.25$, $f=2.0$, system size 76.8$\times$76.8
(grid spacing $h=0.3$). $t=1500.0$.
}
\label{g025f02}
\end{figure}

The presence of a flow term considerably increases the richness of the system
with regard to pattern formation.
Whereas stable structures of the system without
flow can be described in a summarizing fashion as more or less ordered
hexagonal arrays of cells, which may be steady state (with some movement
in the grain boundaries) or oscillatory (with phase shifts of $\approx 2\pi/3$
within a triangle of neighboring cells) or (weakly) turbulent \cite{kassner98},
the system with flow has many more ways of organizing itself.
Figures \ref{g07f04} through \ref{g025f02}  may serve to give a first
impression.
Each of these
figures displays a typical structure for a given temperature gradient
at  a moderate flow.

In Fig.~\ref{g07f04}, the value of the temperature gradient is $\gbar=0.7$,
i.e., larger than $\gbar_c(f=0)$, hence 
the planar interface is destabilized by the flow only. Therefore, no
hexagonall cell structure can develop initially (as long as the pattern is
describable by the linear theory).
We obtain a stripe structure
containing defects, some of which disappear pretty fast, whereas the 
last few  persist for a long time. The final evolution of this pattern up
to several thousand diffusion times will be discussed in II.

Figure \ref{g06f02} is at   $\gbar=0.6$,
where the Mullins-Sekerka instability
is already present. It can be clearly seen that the flow has an organizing
influence on the structure consisting of hexagonal cells: the
dynamical grain boundaries separating differently oriented hexagonal domains
try to orient themselves perpendicular to the flow, so they become
parallel. We did not observe similar ordering of grain boundaries in
simulations without flow, in which
grain boundaries rather tend to form ringlike
structures \cite{kassner98}.


At smaller temperature gradients,  
flows of moderate size 
do not appear to strongly
perturb the basic {\em structure} imposed by the MS instability.
The {\em dynamics} is of course different, since the entire pattern 
drifts against the direction of the flow.
Comparisons with simulations without flow show moreover that structures
display more order (after the same time of dynamical evolution
and starting from the same random initial conditions)
with flow than they do without. As the flow is
increased, new structures develop that we discuss below.

At  $\gbar=0.35$, we find  the flow to promote oscillatory structures.
Figure \ref{g035f02} shows an example of a  {\em topologically ordered}
array of hexagons. Each cell has exactly six neighbors.
The brightnesses of the cells
indicate their different heights; white is high,  black is low.
Different sizes of the cells are due to their being in different phases
of their basic oscillation. There is a phase shift of approximately
$2\pi/3$ between neighboring cells. Phase coherence is not preserved
throughout the entire array of cells as may be noted by comparison of the
lower left and upper right parts of the pictures (not all big bright
cells are exactly the same, and similar statements hold for the
small bright cells and the dark cells). By animated visualization of a
series of pictures, the oscillations are easily verified. Large cells are
seen to become smaller and larger again, periodically. Small cells behave
the same way, the only difference being a phase shift. 
Unfortunately, a movie cannot be transmitted in this media.

{}From earlier
work \cite{kassner98}, these oscillations are known
to occur even without flow; an analytical discussion, not considering
stability issues was given in \cite{daumont97}.
For large grid spacings ($h=0.5$) we saw this dynamical state
already at  $\gbar=0.4$ with our finite-difference code. It was
however observed to  appear later, i.e.,
at smaller  $\gbar$, when the mesh size was reduced, and
we estimated the bifurcation to  $2\pi/3$ oscillations  to happen
between  $\gbar=0.4$ and  $\gbar=0.35$ \cite{kassner98}.
As it turns out, the spectral
code with its higher accuracy does not yet produce the
oscillations at   $\gbar=0.35$, if flow is absent, but it
does so in the presence
of even small flows ($f=1.0$).
In addition, the cellular lattice becomes more
ordered, topological defects are eliminated more efficiently
under flow.

To finish this introductory tour through parameter space, we
consider a temperature gradient of  $\gbar=0.25$ (Fig.~\ref{g025f02}),
which was impossible to do with the finite-difference code because of numerical
instabilities. Even with the spectral code we have to reduce the
mesh size to obtain numerically stable results for $\gbar$ distinctly
smaller than 0.35.

 The example suggests that patterns are
generically weakly turbulent at such a small gradient, i.e., they
show time-dependent nonrelaxational behavior. Nevertheless, 
some ordering influence of the flow may be noted even here
(cells tend to align along the direction perpendicular to the flow)
and becomes much more conspicuous as the flow is increased, to the extent
of rendering  structures regular again at larger flows. We will
return to this question in II.

\subsection{Measured properties}

\begin{figure}[h]
\centerline{ \epsfig{file=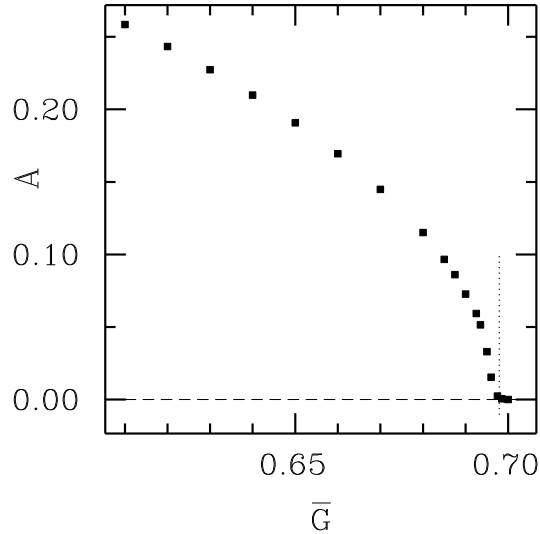,width=8.0cm,angle=0} }
\caption{Amplitude $A$ of steady-state stripe patterns for a fixed flow of
$f=2.0$ as a function of the temperature gradient.
The initial condition was a sinusoidal pattern with wavenumber $0.98$,
with the wave crests
oriented parallel to the $y$ axis. Measurement of the amplitude was
done after $t=1000.0$ in units of the (rescaled) diffusion time. The dotted
line gives the theoretical position of the bifurcation point,
according to Eq.~(\protect\ref{threshold}).}
\label{data_trans}
\end{figure}

In order to characterize the bifurcation from the planar front,
we introduce the standard deviation
$A\equiv\sqrt{\langle(\zeta-\langle\zeta\rangle)^2\rangle}$
as a measure for the amplitude of the
steady state structure, reached in the late stage evolution of an
initially sinusoidal interface. Figure \ref{data_trans} shows
the amplitude so obtained as a function of the temperature gradient.

Due to the small system size (16.0$\times$16.0), it was possible to
keep stripe structures stable well below the threshold of the appearance
of hexagons ($\gbar=2/3$). As expected \cite{hobbs91}, the bifurcation is
supercritical.

In the presence of any nonzero flow, the basic structures
appearing at the instability threshold
are stripes rather than hexagons.
Whereas our large-flow predictions, 
discussed below (and, in more detail, in II),
might require some effort to be realized experimentally,
this result should be of immediate experimental relevance,
since it is valid no matter how small the flow.
Extremely small flows would make the
range of temperature gradients in
which stripes dominate over hexagons very small, but
this range would nevertheless be present and necessarily
observed, at least temporarily,
on crossing the bifurcation threshold via reduction of $\gbar$.
Since the argument for the prevalence of stripes is drawn from
linear stability analysis, it need not continue to hold, once amplitudes
become large enough for nonlinearities to play an important role.
In II, we will see that this indeed happens as long as the flow is not too 
strong.
The transcritical nature of the bifurcation to hexagons will then turn out 
important.

\begin{figure}[h]
\centerline{ \epsfig{file=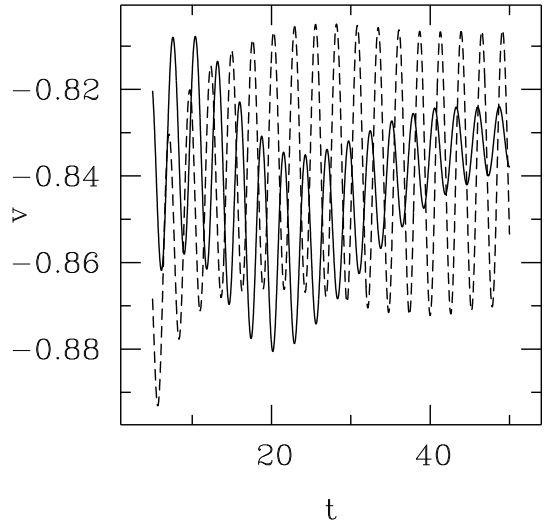,width=8.0cm,angle=0} }
\caption{Oscillations of the lateral pattern velocity after initialization
with a random structure.
Solid line: $\gbar=0.6$, dashed line:  $\gbar=0.7$. For the smaller
value of $\gbar$, the oscillations decay faster.}
\label{homo_osc}
\end{figure}

Measuring the velocity of the interface for values of $\gbar$
which are sufficiently close to the instability threshold we
find that it exhibits damped oscillations. Examples for  $\gbar=0.7$
and $\gbar=0.6$ are presented in Fig.~\ref{homo_osc}. Our discussion
of the velocity measuring procedure in Sec.~III suggests that this
phenomenon may be due to some dynamics superimposed on the drift motion.
This is corroborated by examining the frequency of oscillation.

The figure shows clearly that the frequency is slightly higher for the
larger value of $\gbar$ (for  $\gbar=0.7$,
there are 17 oscillations in the time window
displayed, but only 16 for  $\gbar=0.6$). A precise
determination of the (angular) frequency reveals that it is very close
to the  (angular) frequency $\omega = \sqrt{8k\gbar}$ of {\em homogeneous}
solutions \cite{kassner94} to Eq.~(\ref{basiceq}).
As has been noted before \cite{kassner94}, patterns initiated close to
the threshold $\gbar_c$ oscillate as a whole
 before settling down
into a steady state.

The oscillatory velocity pattern is thus an  effect of the temporal
modulation of the pattern [similar to the $\cos\omega t$ term in
Eq.~(\ref{onedexample1})]
and we should consider the {\em average} over
these oscillations the true velocity.
They are a nongeneric feature of
the current amplitude equation (\ref{basiceq}),
which is not shared by other equations such as the Kuramoto-Sivashinsky
equation. For lower values of $\gbar$, these oscillations are not present.

\begin{figure}[h]
\centerline{ \epsfig{file=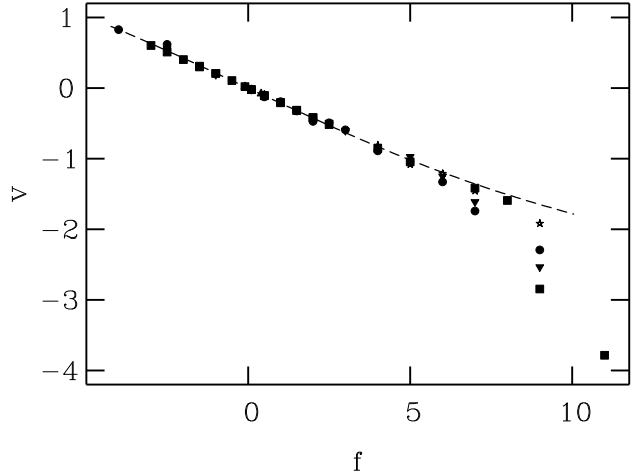,width=10.0cm,angle=0} }
\caption{Measured drift velocity as a function of the flow for
different temperature gradients. Stars:  $\gbar=0.7$,
squares: $\gbar=0.6$, triangles:  $\gbar=0.5$ (most of them covered by
other symbols), inverted triangles: $\gbar=0.4$, circles: $\gbar=0.35$.
Dashed line: Analytic velocity from linear stability analysis for $\gbar=0.6$.
}
\label{vdrift_all}
\end{figure}

Let us now look at the drift velocities measured for different values
of the temperature gradient and the flow. Figure \ref{vdrift_all} collects
some  data corresponding to gradients between  $\gbar=0.35$ and  $\gbar=0.7$.
For flow strengths below $|f| = 6$, all the data points collapse approximately
onto one curve. This is to be expected from the
linear-stability result (\ref{v_asympt_smallf}), which shows that
the $\gbar$ dependence of the drift velocity is weak for small
$f$. The dashed line in the figure shows this result for $\gbar=0.6$.

 On the other hand, the spreading of the data points for $f>6$
does not follow from the corresponding result (\ref{asympt_largef})
for large $f$. Clearly, the linear theory breaks down here. As we
shall see below, the deviation of the drift velocity from the
analytic result happens in the vicinity of the transition to a new pattern,
where the drift velocity is determined by nonlinear effects. We will
discuss this in particular for one temperature gradient,  but there
seems to be a morphology transition in all cases where a strong deviation
from the linear theory arises; the transition is not always to the
same new pattern however. (For $\gbar=0.4$, there seem to be 
two transitions.)

Without flow, the equation of motion (\ref{basiceq}) is symmetric under
the parity transformation $x\to -x$.
Drifting patterns arise, because this symmetry is broken by the
flow term, meaning that we do not have spontaneous symmetry breaking
as, e.g., with parity breaking patterns in the purely diffusive
case \cite{daumont97}. Whereas those patterns are not stable in
extended systems, the present drifting structures are robust.

It is easy to see that flow-induced drifting cells must themselves
be asymmetric with respect to a mirror plane parallel to the $yz$ plane,
even though this asymmetry may be barely perceptible to the eye.
To show this, we
assume the opposite to be true, i.e., $\zeta$ to be symmetric
under  $x\to -x$.
Transforming Eq.~(\ref{basiceq}) to a comoving frame, which results
in $\zeta_t \to \zeta_t-v \zeta_x$, we obtain from the
antisymmetric part of the equation:
\beq
 v \left(2+\kinv+\nu\right) \nabla^2 \zeta_x - f
\frac{\partial}{\partial x} {\cal L}[\zeta] = -2v\zeta_x \nabla^2 \zeta - 2
v \left(\abs{\nabla \zeta}^2\right)_x \>.
\label{basantisym}
\eeq
But this relation must be invariant under an exchange of $x$ and $-x$
{\em and} a replacement of $v$ by $-v$. Hence, the flow term must
vanish, which means it cannot play any role.
This is in contradiction with the fact that
drifting patterns  on large scales are not observed in the absence of flow.

\begin{figure}[h]
\centerline{ \epsfig{file=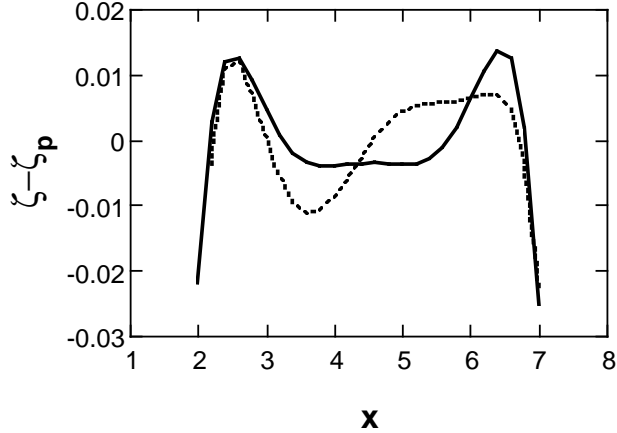,width=8.5cm,angle=0} }
\caption{Deviation of the cell shape from a parabola, for a case
without flow (solid line) and one with a flow $f=2.0$ (dashed line). 
$\gbar=0.5$.
}
\label{asymmetry}
\end{figure}

We have probed the asymmetry of cells in a number of
different ways as a test of our numerical
procedure. The most distinctive method was as follows: first the 2D interface
was cut by a straight line parallel to the $x$ axis through the middle
of a cell, to get its profile. Then the top part of the cell
(everything above a chosen threshold height) was fitted to a parabola.
The difference between the fit function and the actual cell
profile is given for one case without and one with flow in
Fig.~\ref{asymmetry}. Evidently, this difference is symmetric in the former
and asymmetric in the latter case. Since the flow comes from the left
in the picture, we can moreover deduce that the cells are steeper
on the upflow side than on the downflow one.

\section{Conclusions}

The addition of a shear flow to directional solidification near the absolute
stability limit affects the system in several more or less profound ways.

First, the flow breaks the mirror symmetry of the equation of motion. 
This implies that cellular solutions are asymmetric in general. 
Symmetry breaking of this type is known to lead to drifting solutions
in cases where it appears spontaneously \cite{kassner94,daumont97}. 
It produces the same behavior here. The appearance of a drift velocity
can be understood at the level of a linear stability analysis
which moreover provides a decent quantitative estimate for its value.
As expected, this description of the drifting pattern breaks down for
larger flows, where nonlinear effects exert a stronger influence.

Second, the patterns appearing at the instability threshold of the
planar interface for $\gbar>\gbar_c(f=0)$
are not hexagons but stripes. We shall see that this
statement will have to be made more precise in II, because at this moment, we
cannot say anything about the stability of stripe structures.
The planar front becomes unstable via  a supercritical
Hopf bifurcation. Since the transition to hexagons of the system without
flow is transcritical, i.e., hexagons can exist even below the 
threshold of the  diffusive instability, we may expect stripes and hexagons
to interact, at least at small flow strengths. This point will be discussed
in some detail in II.

The main effect of the flow in situations where a structure of hexagonal cells
develops (i.e., for  $\gbar<\gbar_c(f=0)$ and not too large a flow)
is to promote ordering, i.e., the appearance of a hexagonal 
translational lattice. Hence, defects are eliminated more efficiently in 
the laterally moving pattern than in one at rest. Defects that stay
tend to become aligned in the flow to give grain boundaries a preferential
orientation (perpendicular to the flow). Finally, the flow increases 
instability towards local oscillations of cells with relative phase shifts. 
Since these oscillations also act to reduce the number of defects
\cite{kassner98}, the flow reinforces this tendency, again working to
improve translational order.  

\vspace*{0.5cm}
{\bf Acknowledgments}\\

This work was supported by a PROCOPE grant for travel exchanges
by the DAAD (German academic exchange service),
grant no. 9822777, and the Egide
(corresponding French organisation), grant no. 01258ZD.

\appendix

\section{Real space representation of the flow term}

\def\xp{{x^\prime}}
\def\bskip{\mskip-10\thickmuskip}
\def\bskiplrg{\mskip-50\thickmuskip}
\def\bskiph{\mskip-5\thickmuskip}
\def\bskipd{\mskip -3.95\thickmuskip}
\def\bskipt{\mskip -3.45\thickmuskip}
\def\bskips{\mskip -2.8\thickmuskip}
\def\bskipss{\mskip -2.8\thickmuskip}
\def\bskipdm{\mskip -3.6\thickmuskip}
\def\bskiptm{\mskip -3.0\thickmuskip}
\def\bskipsm{\mskip -3.1\thickmuskip}
\def\bskipssm{\mskip -3.1\thickmuskip}
\def\pint{\mathop{\mathchoice{-\bskipdm\int}{-\bskiptm\int}{-\bskipsm\int}{-%
\bskipssm\int}}}
Since the nonlocal term is given by a product in Fourier space
\beq
{\cal L}[\zeta]^\ast \equiv\vert {\bf p}\vert \, \zeta^\ast \>,
\label{ftnonlocast}
\eeq
(where for brevity we denote the Fourier transform by an asterisk)
its position space expression can be obtained as a convolution integral.
Care must be taken,
however, because the inverse Fourier transform of $\vert {\bf p}\vert$ does
not exist.
Therefore, we first rewrite ${\cal L}[\zeta]^\ast$ as a different product.

For a one-dimensional interface, the most convenient approach seems to be
to write
\beq
{\cal L}[\zeta]^\ast =\hbox{sign}(p)  p \, \zeta^\ast  = -\imag
\hbox{sign}(p) (\partial_x\zeta)^\ast\>
\label{oneddecomp}
\eeq
While the (inverse) Fourier transform of the sign function does not exist
as a function,
it is defined in the distribution sense and easily calculable:
\beq
s(x) \equiv \frac1{2\pi} \int_{-\infty}^{\infty} \hbox{e}^{\imag p x}
\hbox{sign}(p) dp
= \frac1{2\pi}\left(\int_{0}^{\infty}  \hbox{e}^{\imag p x} dp -
\int_{0}^{\infty} \hbox{e}^{-\imag p x} dp\right)
= \frac{\imag}{\pi} \frac{\cal P}{x} \>,
\label{s1result}
\eeq
the last expression being the distribution that is pointwise equal to
$\imag/\pi x$
but requires any integral in which it appears to be interpreted as a
principal value.

Hence we obtain for the flow term in 1D:
\beq
-f\partial_x {\cal L}[\zeta] = -\frac{f}{\pi} \,\partial_x
\pint\nolimits_{-\infty}^{\,\infty} d\xp \frac{1}{x-\xp}
\partial_{\xp} \zeta(\xp) \kom
\label{nlocrep}
\eeq
where the bar indicates a principal value
integral. More specifically, we define
\beq
\pint\nolimits_{-\infty}^{\,\infty} d\xp \ldots = \lim_{\eps\to0^+} \left(
\int_{-\infty}^{x-\eps}
d\xp \ldots + \int_{x+\eps}^{\infty} d\xp\ldots \right) \pnt
\label{ppidef}
\eeq

For a two-dimensional interface, another decomposition is more appropriate
\beq
{\cal L}[\zeta]^\ast =-\frac1{\vert{\bf p}\vert} (- {\bf p}^2)  \zeta^\ast
= -\frac1{\vert{\bf p}\vert} (\nabla^2\zeta)^\ast\> \>.
\label{twoddecomp}
\eeq
(In one dimension, the inverse Fourier transform of $1/\vert{\bf p}\vert$
is problematic because of the
divergence at the origin, which in two dimensions is compensated by the
volume element.)
Hence we have 
\beq
{\cal L}[\zeta] = \int_{-\infty}^{\infty} dx' \int_{-\infty}^{\infty} dy'
s({\bf x} - {\bf x'}) {\nabla'}^2 \zeta({\bf x'}) \>,
\label{twodfac}
\eeq
where  $[{\bf p} = (p_x,p_y)]$
\beqa
s({\bf x}) &=& \frac1{4\pi^2}\int_{-\infty}^{\infty}
dp_x\int_{-\infty}^{\infty} dp_y
\frac{-1}{\sqrt{p_x^2+p_y^2}}\, \hbox{e}^{\imag {\bf p x}} \nonumber \\
&=& - \frac1{4\pi^2} \int_0^{2\pi} d\phi \int_{0}^{\infty} dp \frac{p}{p}
\exp\left(\imag p \vert{\bf x}\vert \cos\phi\right) \>.
\label{s2result}
\eeqa
To obtain the second expression, we have oriented the  polar coordinate
system such that
 the ray  $\phi=0$ is parallel to ${\bf x}$
(hence  $ {\bf p x} =  p \vert{\bf x}\vert \cos\phi$).
Therefore,
\beqa
s({\bf x}) &=& -\frac1{4\pi^2} \int_0^{2\pi}  d\phi \left(\frac{\imag{\cal
P}}{\vert{\bf x}\vert \cos\phi}
+ \pi \delta( \vert{\bf x}\vert \cos\phi)\right) \nonumber \\
 &=&- \frac{\imag}{4\pi^2\abs{\bf x}} \pint\nolimits_{0}^{2\pi} d\phi
\frac1{\cos\phi}
-\frac{1}{4\pi} \int_0^{2\pi} d\phi \frac{1}{\vert{\bf x}\vert \vert
\sin\phi\vert} \left[\delta\left(\phi-\frac{\pi}2\right) +
\delta\left(\phi-\frac{3\pi}2\right) \right] \nonumber \\
&=& - \frac1{2\pi\vert{\bf x}\vert} \>.
\label{s2result2}
\eeqa
The principal value integral of $1/\cos\phi$ vanishes as it
extends over an entire period. Thus we arrive at
the following final expression for the flow term
\beq
-f\partial_x {\cal L}[\zeta] = \frac{f}{2\pi} \,
\partial_x  \int_{-\infty}^{\infty} dx' \int_{-\infty}^{\infty} dy'
\frac1{\vert {\bf x-x'}\vert }
{\nabla'}^2 \zeta({\bf x'}) \>.
\label{nlocrep2}
\eeq

Both the 1D and 2D expressions clearly exhibit the nonlocality of the flow
term and its odd parity
symmetry. 
All the other terms in Eq.~(\ref{basiceq}) have
even parity, so the flow term provides a symmetry breaking mechanism.

Note also that the nonlocal kernels appearing in these expression are
almost the simplest possible,
if one thinks in terms of gradient expansions. Local terms
produce successive powers
of derivatives, corresponding to powers of ${\bf p}$ in Fourier space.
Nonlocal terms would
be connected with powers of $1/\vert{\bf p}\vert$, which must be
compensated for (in order to
keep things finite at small ${\bf p}$) by spatial derivatives.

\section{Analytic calculation of drift velocities}

In order to solve Eqs.~(\ref{eqomegar}) and (\ref{eqq}) analytically, we
first cast them in
a simpler form introducing
\beq
\tilomr = \omega_r+2q^2 \>.
\label{deftomrtil}
\eeq
This yields
\beqa
\tilomr^4 + \tilomr^2\left(8\gbar-q^4-8q^2\right) - \frac{f^2 q^4}{4} &=& 0 \>,
\label{eqomtilr} \\
f^2 q^2 \left(\tilomr-2q^2\right) -2\tilomr^3\left(4\tilomr-2q^2-8\right)
&=& 0 \>.
\label{eqtilq}
\eeqa
We then consider the limiting cases $f\ll1$ and $f\gg1$.
The first of these is very simple. As we know that there are solutions
without flow,
we set, as a first approximation, $f=0$ in Eqs.~(\ref{eqomtilr}) and
(\ref{eqtilq}).
$\omega_i$ will still remain $f$ dependent via (\ref{omegai}), which takes the
form
\beq
\omega_i = \frac{fq^2}{2\tilomr} \>.
\label{newomegai}
\eeq
We get immediately
\beq
\tilomr = 2+\frac12 q^2 \label{solomrtsmf}
\eeq
and the Eq.~(\ref{eqomtilr}) for $q$ can be reduced to quadratic.
The result is
\beqa
q^2 &=& -4 + \frac4{\sqrt{3}}\sqrt{4+2\gbar} \>, \\
\omega_r  &=& -2q^2 +  \frac2{\sqrt{3}}\sqrt{4+2\gbar}  \>, \\
\omega_i &=& f \left(1 -
\frac{\sqrt{3}}{\sqrt{4+2\gbar\vphantom{\bar{\gbar}}}} \right)  \>,
\label{omi_asympt_smallf}\\
v = -\frac{\omega_i}{q}  &=& -\frac{f
\sqrt{3}}{2\sqrt{4+2\gbar\vphantom{\bar{\gbar}}}}
\left(\sqrt{\frac{4+2\gbar}{3}}-1\right)^{1/2} \>.
\label{v_asympt_smallf}
\eeqa
As an example, for $\gbar=0.7$, this gives a drift velocity $v=0.218 f$, and
the dependence on $\gbar$ is weak. Note that for this $\gbar$ value
$\omega_r<0$ and hence the structure will decay to a planar front
(moving along as it does so at the the calculated velocity).

The case $f\gg 1$ is slightly more complicated. Considering (\ref{eqomtilr})
and the fact that $\gbar$ is of order 1 in the interesting parameter
range, we see (from the signs in the equation) that the only possible
dominant balance for a {\em pair} of terms is
\beq
\tilomr^4 \sim \frac{f^2 q^4}{4} \>. \label{dombal1}
\eeq
Inserting this in (\ref{eqtilq}) and combining the two equations, we
find however
\beq
\tilomr \sim 2 q^2 \>, \label{dombal2}
\eeq
which is in contradiction with (\ref{dombal1}) for $f\gg1$
and $q=O(1)$. This implies
that dominant balances in (\ref{eqomtilr}) must contain {\em three}
 terms at least
and that the wavenumber $q$ of the fastest-growing mode has to scale, too.
It must increase with $f$ and
Eqs.~(\ref{dombal1}) and (\ref{dombal2}) suggest the scaling $f\sim q^2$.
Hence, we set
\beq
f = a q^2 \>, \label{fscal}
\eeq
assuming $a= O(1)$. Inserting this into  Eqs.~(\ref{eqomtilr}) and
 (\ref{eqtilq}) and using $q\gg1$, we arrive at
\beqa
\tilomr^4 - \tilomr^2 q^4 - \frac{a^2q^8}{4} &=& 0 \>, \label{tilomlargef}\\
a^2 q^6 (\tilomr^2-2q^2) &=& 8\tilomr^4 - 4\tilomr^3q^2 \>. \label{qlargef}
\eeqa
Equation (\ref{tilomlargef}) can be solved for $\tilomr^2$:
\beq
\tilomr^2 = \frac{q^4}{2}\left(1+\sqrt{1+a^2}\right) \>.
 \label{restilomrlargef}
\eeq
Using this  in (\ref{qlargef}), we end up, after some simplifications,
with a relation determining $a$:
\beq
\frac{\sqrt{2}}{8} \left(1+\sqrt{1+a^2}\right)^{3/2} - \sqrt{1+a^2} = 0 \>.
\label{aeq}
\eeq
The numerical solution of this algebraic equation yields $a=28.88$. 
(There is only one real solution.)
We then obtain
\beqa
q &=& \sqrt{\frac{f}{a}} \approx 0.186 \sqrt{f} \>, \\
\omega_r &=& \left(\sqrt{\frac{1+\sqrt{1+a^2}}{2}}-2\right) \frac{f}{a} \approx
0.065 f \>, \\
\omega_i &=& \frac{2f}{\sqrt{2\left(1+\sqrt{1+a^2}\right)}} \approx 0.129 f
\>,\label{omi_asympt_largef} \\
v &=& -\frac{\omega_i}{q} \approx -0.695 \sqrt{f} \>.
\label{asympt_largef}
\eeqa
All of these expressions are independent of $\gbar$, as they must,
being leading order results for $f\gg\gbar$.


\begin{thebibliography}{10}
\bibitem{footnote1}
In \protect\cite{kassner98}, structures were simulated up to
250$\,$000 rescaled diffusion times $\tilde\tau$.
For typical solids, $\tilde\tau=\tau/\epsilon$ with $\tau=4D/V^2\approx 1000$s
($D\approx 10^{-5}$cm$^2$/s, $V=2\mu$m/s), whereas for liquid crystals,
$\tau\approx 10^{-3}$s is achievable 
($D\approx 10^{-7}$cm$^2$/s, $V=100\mu$m/s).
\end{thebibliography}
\end{document}